# Guidelines for releasing a variant effect predictor

Benjamin J. Livesey[1], Mihaly Badonyi[1], Mafalda Dias[2], Jonathan Frazer[2], Sushant Kumar[3], Kresten Lindorff-Larsen[4], David M. McCandlish[5], Rose Orenbuch[6], Courtney A. Shearer[6], Lara Muffley[7], Julia Foreman[8], Andrew M. Glazer[9], Ben Lehner[10], Debora S. Marks[6,11], Frederick P. Roth[12], Alan F. Rubin[13], Lea M. Starita[7] and Joseph A. Marsh[1]*

[1]*MRC Human Genetics Unit, Institute of Genetics and Cancer, University of Edinburgh, Edinburgh, UK*
[2]*Centre for Genomic Regulation (CRG), The Barcelona Institute of Science and Technology, Barcelona, Spain*
[3]*Department of Medical Biophysics, University of Toronto; Princess Margaret Cancer Centre, University Health Network, Toronto, Ontario, Canada*
[4]*Linderstrøm-Lang Centre for Protein Science, Department of Biology, University of Copenhagen, Copenhagen, Denmark*
[5]*Simons Center for Quantitative Biology, Cold Spring Harbor Laboratory, Cold Spring Harbor, NY, USA*
[6]*Department of Systems Biology, Harvard Medical School, Boston, MA, USA*
[7]*Department of Genome Sciences, University of Washington and the Brotman Baty Institute for Precision Medicine, Seattle, WA, USA*
[8]*European Molecular Biology Laboratory, European Bioinformatics Institute, Wellcome Genome Campus, Hinxton, Cambridge, UK*
[9]*Vanderbilt University Medical Center, Nashville, TN, USA*
[10]*Wellcome Sanger Institute, Cambridge, UK; Universitat Pompeu Fabra (UPF), Barcelona, Spain; Institució Catalana de Recerca i Estudis Avançats (ICREA), Barcelona, Spain*
[11]*Broad Institute of MIT and Harvard, Boston, MA, USA*
[12]*Department of Computational and Systems Biology, University of Pittsburgh School of Medicine, Pittsburgh, PA, USA*
[13]*Bioinformatics Division, Walter and Eliza Hall Institute of Medical Research; Department of Medical Biology, University of Melbourne, Parkville, Australia*

*joseph.marsh@ed.ac.uk*

## Abstract

Computational methods for assessing the likely impacts of mutations, known as variant effect predictors (VEPs), are widely used in the assessment and interpretation of human genetic variation, as well as in other applications like protein engineering. Many different VEPs have been released to date, and there is tremendous variability in their underlying algorithms and outputs, and in the ways in which the methodologies and predictions are shared. This leads to considerable challenges for end users in knowing which VEPs to use and how to use them. Here, to address these issues, we provide guidelines and recommendations for the release of novel VEPs. Emphasising open-source availability, transparent methodologies, clear variant effect score interpretations, standardised scales, accessible predictions, and rigorous training data disclosure, we aim to improve the usability and interpretability of VEPs, and promote their integration into analysis and evaluation pipelines. We also provide a large, categorised list of currently available VEPs, aiming to facilitate the discovery and encourage the usage of novel methods within the scientific community.

## Background

Many different computational methods, known as variant effect predictors (VEPs), have been developed to assess the likely impacts of genetic variants [1–3]. These tools are often applied in the analysis and interpretation of human genetic variation, and are widely used in clinical genetics investigations for variant prioritisation and assessing potential pathogenicity. Importantly, however, the extent to which their predictions can be relied upon for making genetic diagnoses is still limited [4]. VEPs also show considerable utility in other non-human-specific applications, such as evolutionary analyses [5,6] and protein engineering [7,8].

VEPs are highly heterogeneous in terms of their underlying algorithms, the data upon which they are based, the meaning and interpretation of their predictions, the format in which their predictions are shared, and the way in which the methods themselves are made available. While substantial progress is being made in the VEP field, this diversity of methods also makes it difficult for end users to make an informed choice as to the best VEP for their purpose. There is also a particular problem related to unbiased assessment, whereby nearly every new predictor that is released claims to show superior performance to all methods against which it is compared [9]. Consequently, there has been increasing focus on independent benchmarking of VEP methodologies in recent years [10–13]. However, identifying new methods can be difficult, given their large number, the fact that they are often disparately named in the literature (*e.g.* predictors of 'variant effect', 'functional effect', 'deleteriousness', 'pathogenicity', 'mutational impact', etc, or simply just 'missense predictors'). In addition, great effort is often required to obtain their predictions. Finally, fair evaluation of VEP performance requires detailed knowledge of the data used to train different methods, which is often unclear from the original publications.

The Atlas of Variant Effects (AVE) Alliance coordinates researchers from around the world who are working on both computational and experimental approaches for characterising variant effects on a large scale, including VEPs and high-throughput experimental strategies, termed Multiplexed Assays of Variant Effect (MAVEs) [14]. The AVE "Analysis, Modeling, and Prediction" workstream is spearheading efforts to coordinate, develop and assess computational methods for variant effect prediction and for the analysis of MAVE datasets. Based on our extensive experience in the development and use of computational tools for variant analysis, here, we provide guidelines and recommendations that we believe should be considered when releasing a novel variant effect prediction method (Figure 1). Our scope here is primarily on VEPs that score pathogenicity, evolutionary tolerance or general impacts on fitness or function. However, we recognise that there are many tools for predicting other aspects of variant effects, such as changes in biophysical properties like protein stability [15], binding affinity [16] and aggregation propensity [17], or alterations in splicing [18]. We believe that most of our recommendations will be applicable to these tools as well. While some of our advice is specific to predictors of protein variant effects, we also discuss issues relating to nucleotide-level and non-coding predictors.

We hope that these guidelines will improve the evaluation of novel methods, and facilitate their seamless incorporation into existing analysis pipelines. Furthermore, we believe that this will contribute to the broader adoption and utilisation of VEPs within the scientific community, thereby accelerating our understanding of genetics and improving patient care. Ultimately, our goal is to support the creation of tools that are both scientifically rigorous and widely accessible, paving the way for advances in personalised medicine and genetic research.

## Results

### *Sharing of methods and code*

Although many users may only be interested in the outputs of VEPs, *i.e.* the variant effect scores, it is also imperative that the methods themselves be made available. This allows novel variants to be tested and for methods to be more comprehensively evaluated. We strongly suggest that variant effect prediction methods be made freely available and open source, with a clear Open Source Initiative (https://opensource.org) approved licence. Open-source practices not only catalyse widespread adoption and application, but also underpin the collaborative ethos of the scientific community. By making VEP methods and their corresponding codebases accessible and clearly documented, developers empower researchers across the globe to contribute to the evolution of these tools, enhancing their accuracy, efficiency, and utility. Furthermore, open-source licensing fosters an environment where innovations can be rapidly disseminated and iteratively improved upon. As we move into an era where computational methods are likely to be given an increasingly important role in making genetic diagnoses, it is essential that these methodologies be as open as possible, to ensure that their outputs can be trusted. Making code available also makes it possible to calculate scores using, for example, different reference genomes, with potential importance in health equity [19].

In the past, many VEP methodologies have been made available as web servers, whereby individual variants can be queried. While this can be convenient for end users who are interested in small numbers of variants, making a method available only as a web server severely limits the potential for a method to be independently assessed. At the very least, web servers should offer an application programming interface (API) for bulk queries, if a pre-calculated download is not an option. Concerningly, we have found many examples of such online predictors disappearing from the web after only a few years, undermining their long-term utility.

Hosting the code for a VEP on a public, open-source platform like GitHub (https://github.com) or Huggingface (https://huggingface.co) provides high levels of visibility, easy version control, and the opportunity to easily integrate documentation. Repositories such as Kipoi [20] are also useful for depositing models, facilitating broader access to the necessary tools for exact replication of model predictions. Releasing models with their trained parameters is crucial for reproducibility and utility. This practice addresses the inherent stochasticity in training machine learning models, ensuring that the reproducibility of a VEP is not compromised. Releases should also be stably archived using a Digital Object Identifier (DOI) providing service, ensuring reproducibility even when the model is updated. End users can also easily modify open-source code to meet their own unique requirements such as integration into specific prediction pipelines and processes. A containerised version of the method, utilising platforms such as Docker (https://www.docker.com) or Apptainer (formerly known as Singularity, https://apptainer.org), can also be very useful, especially for cross-platform analysis or where installation poses challenges. These tools encapsulate the method and its dependencies in a container, ensuring that it can be run seamlessly across different computing environments.

It is also important to clearly document the methodology underlying a novel VEP. This should include a clear list of all the features included in the final model with links to sources and code or replicable methodology that can be used to engineer these features if necessary. For methods including three-dimensional structural data, the source of each structure should be clearly identified. Ideally, whenever licensing permits, providing direct access to source files ensures reliability and reproducibility by avoiding dependence on external databases.

In addition to making VEP methodologies and resources transparently available, it is helpful to communicate the computational cost and runtime associated with these tools. A VEP capable of running genome-wide analyses on a standard laptop offers different possibilities compared to one requiring substantial computational resources for only a few protein assessments. This distinction not only impacts the practicality of the tool for various research applications, but also raises important considerations regarding energy consumption and sustainability [21].

## *Interpretability of variant effect scores*

The outputs produced by different VEPs can vary widely. For tools that predict effects on specific biophysical properties like stability or interactions, the meaning of the outputs is often very clear (*e.g.*, predicted ΔΔG in units of kcal/mol). However, most VEPs provide a variant effect score that may be interpreted as being related to the likelihood of a given variant being pathogenic, or damaging to function or fitness, without any consideration of mechanism. The interpretation of these scores is often difficult and the scales can vary widely. The most common scale ranges from zero to one, with zero being the least and one the most damaging. However, we note that this directionality is opposite to what is commonly used for the functional scores that are the outputs of MAVE experiments, in which a value of one often represents wild-type fitness and zero corresponds to the fitness of a null (*e.g.*, nonsense) variant [22]. While, ideally, VEPs and MAVEs would be calibrated to similar scales, we suggest that creators of new VEPs consider adopting zero-to-one scales of least-to-most damaging, so long as this does not obfuscate the interpretation of the variant effect score. This matches the most common convention and aligns with the directionality used by the large majority of current methods. Standardising variant effect scales will not only simplify interpretation and comparison for non-experts but also reduce the potential for human error during data analysis.

It is important to include an explanation of how scores can be compared. For many methods, variant effect scores can be compared across genes (*e.g.,* two different variants with the same score from two different genes would be considered equivalent in terms of their likelihood of being pathogenic). However, for others, the scales are defined at the level of individual genes, and scores for variants from different genes are not necessarily comparable. For example, DeepSequence models are generated on a per-protein basis, with the scores representing the likelihood ratio between mutant and wild-type residues [23]; thus scores from different proteins are not directly comparable.

Some methods provide labels along with variant effect scores. These are often desired by end users, but also come with a risk of overinterpretation. The rationale and thresholds must be clearly explained and justified, and care should be taken about the choice of labels. For example, AlphaMissense classifies many possible human variants as *'likely pathogenic'* and *'likely benign'* [24]. However, there already exist very clear clinical definitions of these terms that are completely distinct from the definitions used by AlphaMissense [4,25]. This has considerable potential to confuse end users, who may include patients or patient families, if these labels are applied to variants of uncertain significance. If classifications are to be provided alongside variant effect scores, we suggest that terms that are distinct from the clinical labels be used. For example, the widely used PolyPhen-2 predictor defines thresholds for *'possibly damaging'* and *'probably damaging'* [26]; these terms should have a much lower chance for confusion with the well-established clinical calssifications. An alternative could be to use the Sequence Ontology terms *'functional_normal'* and *'functionally_abnormal'* [27]. As more mechanism-centric predictive methods are introduced, developers could consider using scores and thresholds specific to different classifications associated with the '*functionally_abnormal*' term.

One emerging strategy for facilitating the use of VEP scores as evidence in clinical variant interpretation is through calibration to ACMG evidence strength levels [28,29]. Importantly, however, even after using a well-validated calibration, references to pathogenicity should only describe scores as *evidence towards* pathogenicity or benignity, rather than defining variants as such.

## *Accessibility of predictions*

The success of a VEP is intricately linked to the availability of its outputs. The free and unrestricted availability of these scores is essential for the method to be widely used. Ensuring that these data are not only available but also, Findable, Accessible, Interoperable, and Reusable aligns with the FAIR Guiding Principles for scientific data management [30]. Adhering to FAIR principles in disseminating variant effect scores facilitates broader participation in genomic research, enhances the reproducibility of scientific findings, and accelerates the translation of genomic data into actionable clinical insights.

Unfortunately, certain methods impose restrictive licensing terms on their predictions, hindering independent performance assessments and, consequently, limiting user confidence and impeding integration into clinical variant assessment frameworks. We therefore advocate for freely available data to enable scientific discovery and clinical decision-making. The argument has been made against making variant effect scores freely available to avoid their incorporation into other predictors and thus confounding performance assessment [31]. We agree that there are potential complications arising from such ensemble or meta-predictors, as discussed below. However, we believe that the issue of restricted data preventing the very assessments needed to address potential confounding effects is far more concerning, and that such closed methods can never receive the open, independent assessments needed to be considered trustworthy by the community.

The methodology behind a VEP dictates the most appropriate format for sharing its predictions. For many currently available VEPs, predictions are performed at the protein level. Thus, scores should be provided with respect to the appropriate reference sequence against which the prediction was performed. In our experience, most protein-level VEPs output predictions using canonical UniProt protein sequences. Going forward, we recommend that developers utilise transcripts recommended by the Matched Annotation from NCBI and EMBL-EBI (MANE) collaboration [32]. The MANE Select transcript set includes a default recommended transcript for nearly all protein coding genes and matches the UniProt canonical isoform in the vast majority of cases. In addition, the MANE Plus Clinical transcripts are defined for the relatively small number of genes where a single transcript is not sufficient to report all clinically relevant variants. Therefore, we suggest that, for protein-centric-methods, variant effect scores ideally be provided for all possible single amino acid substitutions across all protein sequences corresponding to MANE Select and MANE Plus Clinical transcripts. However, we recognise that this is not always computationally feasible. In these cases, we suggest that predictions be provided for as many human proteins as possible, focusing on those of greatest clinical relevance (e.g., genes included in Gene Curation Coalition database [33] or the ACMG secondary findings list [34,35], and those for which MAVE datasets have been published, enabling MAVE-based benchmarking).

Other VEPs make predictions at the nucleotide level. The further advancement of such methods is critical to interpreting the vast majority of human genetic variation that occurs in non-coding regions [36]. For methods that make predictions outside of exonic regions, variant effect scores should be shared using genomic coordinates referenced against a versioned reference genome assembly.

In some cases, protein-based methods have their predictions shared in terms of genomic coordinates. While this has some advantages in terms of facile incorporation into genomic analysis pipelines, we suggest that, if predictions are made at the protein level, then predictions should also be provided at the level of the same protein sequences. In addition, most single amino acid substitutions cannot be achieved by single nucleotide changes, thus losing some information if only nucleotide-level predictions are provided. While this has no impact on analyses of single nucleotide variants, there are many examples of pathogenic single amino acid substitutions caused by multi-nucleotide changes. These substitutions may also be of interest for other reasons, such as comparison to MAVEs or for protein engineering applications. Separate tools, such as the Ensembl Variant Effect Predictor [37], or the Ensembl REST API [38] and EMBL-EBI Proteins API [39], can be used to map protein-level variants to genomic coordinates, if necessary. When predictions are performed or provided at the nucleotide level, but analyses are at the protein level, there can be some ambiguity if different variant effect scores are provided for different single nucleotide variants that translate into the same amino acid change. We suggest reporting the most deleterious score, in addition to also sharing the nucleotide-level predictions.

Some VEPs are able to make predictions for variants other than single amino acid or single nucleotide substitutions. At the protein level, it may be possible to provide comprehensive predictions across the human proteome for truncations and for single amino acid insertions and deletions. However, it would be unrealistic to provide predictions for all possible variants when considering larger sequence changes involving indels and multi-amino acid substitutions. Similarly, for nucleotide-level predictors, it may be infeasible to provide complete predictions for anything other than single nucleotide variants for a limited subset of the genome. In these cases, the availability of the method for users to run specific predictions of interest is absolutely essential. In addition, predictions could be specifically provided for larger sequence variants known to be pathogenic [40] or present in the human population [41].

When sharing variant effect scores for single amino acid substitutions clearly mapped to stable sequence identifiers, using a simple, common format (*e.g.,* P316D) should be fine, although using the Human Genome Variation Society (HGVS) notation [42] (*e.g.,* p.Pro316Asp) would slightly reduce ambiguity in terms of possible confusion with nucleotide substitutions. For larger and more complex variants, we recommend considering the Global Alliance for Genomics and Health (GA4GH) Variation Representation Specification (VRS) [43]. By adhering to the GA4GH VRS, researchers and clinicians can ensure that complex variant data are not only accurately captured but also remain interpretable and usable across various genomic research and clinical diagnostic applications.

Although most of the current interest in VEPs is focused on human genetic variation, and many VEPs have been developed that only provide predictions for human variants, some VEPs, particularly those based on unsupervised learning approaches, are applicable to variants from any species. While it is clearly not realistic to provide predictions for all variants across all species, we suggest that, in addition to predictions across the human proteome, variant effect scores be provided for any variants present in MaveDB [44] and/or ProteinGym [45] to facilitate independent benchmarking and analysis.

For sharing variant effect scores, as well as other essential data discussed here, we strongly recommend deposition in a well-established public repository that provides a DOI for reference, such as Zenodo ([https://zenodo.org](https://zenodo.org)), Dryad ([https://datadryad.org](https://datadryad.org)), or the Open Science Framework ([https://osf.io](https://osf.io)), rather than hosting them on the authors' website. This practice not only ensures the long-term availability and utility of the data but also helps its distribution, since many of these repositories have an API that allow fast programmatic access to data.

### *Availability of training data*

Most VEPs that have been developed to date are based on supervised learning strategies based on training against labelled datasets of variants, usually split into pathogenic and benign, sourced from databases like ClinVar [40] and gnomAD [41]. A critical issue in the field of variant effect prediction is that of data circularity, whereby the performance of VEPs is evaluated using either variants that were directly or indirectly used in training, thus inflating apparent performance [46]. Related to this, the performance of different VEPs is heavily influenced by the test datasets, and many tools perform markedly worse when applied to novel missense variants [11].

To address this problem, recent studies have used correlations with independent MAVE datasets to compare VEP performance [10,12,45]. While this can be useful to compare different VEPs, it is worth noting that MAVEs do not always probe functions that are central to the development of disease or use a disease-relevant tissue context. If a more traditional assessment of discrimination between pathogenic and benign variants is desired, it is essential to ensure that none of the variants used in VEP training, or other variants at the same positions, are present in the evaluation set, to avoid confounding from type 1 circularity [46]. Moreover, given the issues associated with type 2 circularity, in which a trained model tends to overcall all variants in a gene as pathogenic or benign [46], it would be safest to exclude from evaluation any variants from genes used in training of the VEP, or even genes homologous to any genes used in training. Note that this does not preclude the development of gene-specific VEPs using supervised approaches, as type 2 circularity would only be relevant for comparisons involving variants from multiple genes. This reiterates the importance of making source code and models freely available, to allow performance when using different sets of training data to be assessed.

Given these issues with circularity, it is crucial for the integrity and transparency of a VEP that all variants employed in its training are disclosed upon release. Ideally, these should be shared in the same format as the variant effect scores themselves, rather than merely referencing the databases, due to the dynamic nature of these resources and the potential variability in mapping methods to different sequence identifiers. In situations where controlled access datasets are used and a comprehensive list of training variants cannot be openly shared, it becomes imperative to explicitly detail the version of the dataset, along with the processing and filtering methods applied. This ensures that, despite the restrictions, the original training set can be accurately reconstructed by others. Consequently, we strongly advise against using any private or commercial datasets for training if the variants cannot be fully disclosed.

Difficulties associated with circularity can become particularly acute with ensemble or meta-predictors, which use the outputs of other VEPs as features in their training. If other supervised models are used as features, then the identities of the variants used to train those models are required for fair assessment. Thus, when developing a novel meta-predictor, we strongly recommend including only the scores from unsupervised VEPs as features. If scores from supervised VEPs are included, then one should ensure that the identities of all variants used to train those models are also available, and that this is taken into consideration during testing and validation.

Some VEPs have been released that do not train on pathogenic variants, but do contain information on the allele frequencies of variants present in the human population, or their frequencies in primate species [24,47]. It is essential that the identities of these variants, or their mapped human variants be provided. We emphasise that such VEPs face the same issues of circularity in performance evaluation as other supervised VEPs. In particular, allele frequency is very commonly used as evidence in the classification of variants as benign [4]. Thus, when assessing discrimination between pathogenic and benign variants (*e.g.*, using variants from ClinVar), a VEP that is trained or tuned using allele frequencies will have effectively been exposed to much of the benign dataset, which can inflate apparent performance [48].

An important application of VEPs is in the interpretation of extremely rare variants. As it has been shown that common benign variants are not representative of rare benign variants [49,50], users may wish to choose VEPs that perform well on test sets of exclusively rare variants. Consequently, those training VEPs may wish to consider excluding common benign variants from their training sets or downweighting their influence.

An increasing number of VEPs are also now directly including MAVE data in their training [50–52]. This introduces new circularity issues, and can confound MAVE-based benchmarking attempts if the datasets used for training are not excluded. As long as the MAVE datasets are present in MaveDB or a benchmark such as ProteinGym, it should be sufficient to cite their accession if used in training. In the event that MAVE data is hosted at a location that may become unavailable (e.g., on a group's website), then all variants used for training should be provided, similarly to database-sourced training variants.

There are unique issues associated with VEPs that work on the nucleotide level and are focused on predicting non-coding variant effects. It is crucial for these models to specify the resolution used in training, the genomic regions used (*e.g.,* whole genome, promoters, UTRs etc.) and the molecular/evolutionary modalities considered. These details directly influence how the effects of variants are interpreted and delineate the scope of sequences for which the model can accurately provide predictions.

Increasingly, many state-of-the-art VEPs are based upon unsupervised approaches, often taking multiple sequence alignments as input [53,54]. Although it has not been common practice in the past, we suggest that it is important to make the sequence alignments underlying these models available, along with careful documentation of how the alignments were generated. This would allow assessment of the extent to which the alignment depth and quality influence prediction performance. Furthermore, non-human variants, especially from primates, have occasionally been used as 'benign' variants for VEP evaluation. This could lead to another level of circularity, if these non-human species have been included in the sequence alignment. Thus the availability of sequence alignments and knowledge of the species on which the model is based, can be crucial.

The other increasingly popular unsupervised approach, protein language models, are trained directly on protein sequence information and do not require alignment generation for prediction [55]. While the identity of the databases used to train such models is often provided, model-specific clustering and filtering procedures can obfuscate the exact sequences that were used during training. We suggest that authors of language models and similar methods provide both the database version and all sequence identifiers that went into training the final version of the model.

### *A list of currently available variant effect predictors*

To increase the visibility and discoverability of new VEPs, we have compiled an extensive list of tools at https://www.varianteffect.org/veps. This includes our own classifications in terms of their underlying methodologies and features, based on descriptions provided in the original publications. We also include details on the author-recommended pathogenicity prediction thresholds, and links to their web servers, variant effect scores, training data and code downloads. A current snapshot of our VEP list is provided as Table S1. While this list is not yet fully comprehensive, given the huge number of tools that have been published, we are actively adding new methods as we identify them, and we strongly encourage suggestions for new methods to be included using the web form available at that site. Furthermore, while the current list is heavily weighted by those methods developed for and applied to the problem of predicting pathogenicity in single amino acid substitutions, our aim is to eventually cover all aspects of variant effect prediction.

## Conclusions

The guidelines presented herein aim to streamline the development, sharing, and evaluation of VEPs, addressing key issues such as data availability, interpretability, method transparency, and data circularity. By advocating for the free dissemination of variant effect scores, open-source sharing of methods and code, and rigorous standards for training data, we hope to enhance the reliability, usability, and scientific integrity of VEPs within the community. The promotion of best practices in sharing both predictions and methodologies not only facilitates independent assessment and integration into clinical frameworks, but also fosters collaboration and innovation in the field. As VEPs improve, they are likely to be given greater weight in the clinical interpretation of genetic variants, either alone [29], or in combination with increasingly available MAVE data [56]. Ultimately, adhering to these guidelines will significantly contribute to the advancement of personalised medicine and our understanding of the genetic basis of disease, marking a step forward in the collective effort to harness computational predictions for clinical and research applications.

## Funding

MD and J Frazer are supported by the Spanish Ministry of Science and Innovation (PID2022-140793NA-I00). KLL received funding from the Novo Nordisk Foundation (NNF18OC0033950). DMM received funding from the National Institutes of Health (NIH) (R35GM133613). RO and DSM are supported by the Chan Zuckerberg Initiative Neurodegeneration Challenge Network (CZI2018-191853). LM and LMS received funding from the NIH (RM1HG010461). J Foreman received funding from the Wellcome Trust (WT223718/Z/21/Z). AMG received funding from the NIH (R35GM150465). BL and MB received funding from the European Research Council (ERC) under the European Union's Horizon 2020 research and innovation programme (grant agreement No. 883742). DSM received funding from the NIH (1R01CA260415). AFR received funding from the NIH (UM1HG011969, RM1HG010461, R01HG013025) and grant funding from the Australian Government. JAM is supported by the ERC (grant agreement No. 101001169) and by the Medical Research Council (MRC) Human Genetics Unit core grant (MC_UU_00035/9).

## Acknowledgments

We thank Sarah Hunt for helpful comments on the manuscript and Alex Hopkins for administrative support.

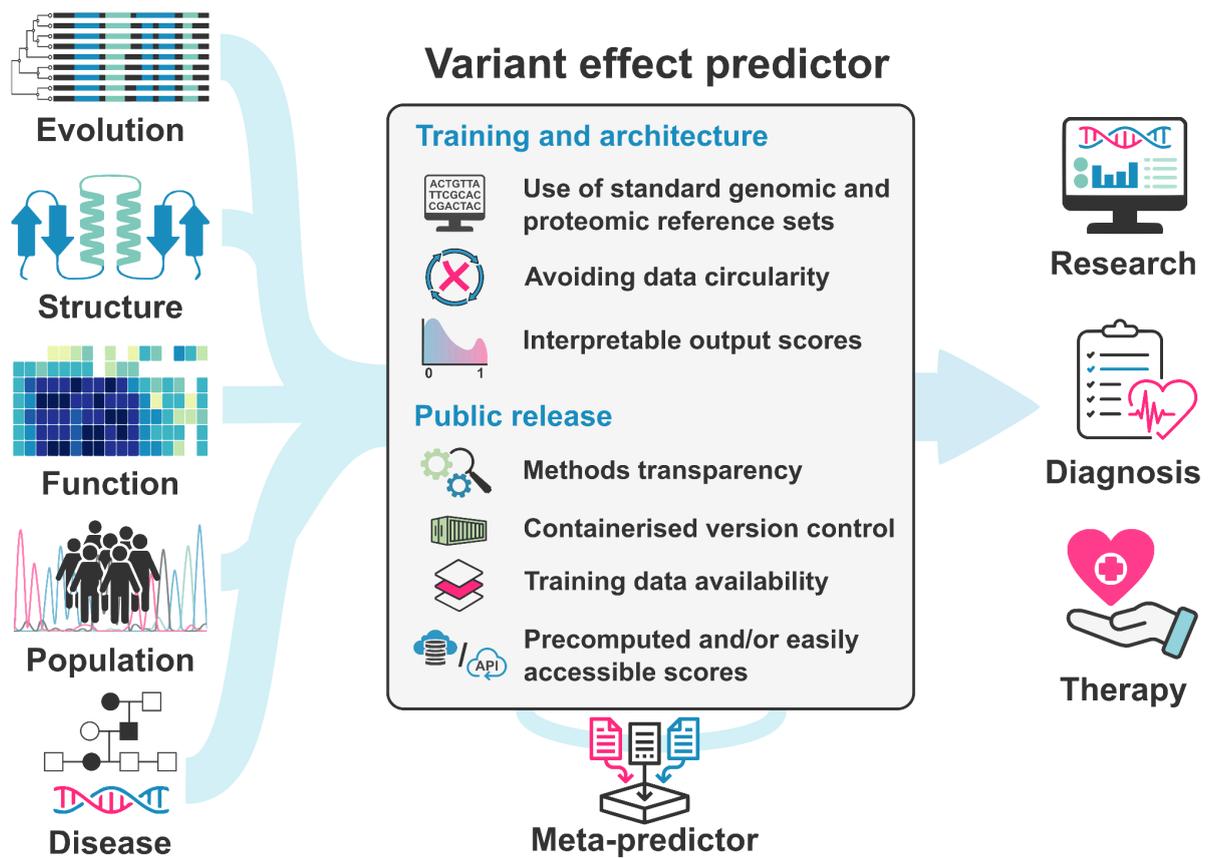

**Figure 1:** Overview of variant effect predictors, including common inputs and outputs, and guidelines for development and release.